\documentclass[
 aps, pra, superscriptaddress, amsmath,amssymb,
 twocolumn,
floatfix,
longbibliography,10pt
]{revtex4-2}

\usepackage[final]{graphicx}
\usepackage{dcolumn}
\usepackage{bm}
\usepackage{hyperref}
\usepackage[utf8]{inputenc}
\usepackage{epsfig}
\usepackage{braket}
\usepackage[capitalize]{cleveref}
\crefname{section}{Sec.}{Secs.}
\crefname{equation}{Eq.}{Eqs.}
\crefname{figure}{Fig.}{Figs.}
\usepackage{enumerate}
\usepackage{color}
\usepackage{graphicx}
\usepackage{tikz}
\usetikzlibrary{quantikz2,calc,shadings}
\usepackage{appendix}
\usepackage{nicematrix}
\usepackage{booktabs}      
\usepackage{bbm}
\usepackage{comment}
\usepackage{makecell}

\usepackage{enumitem}
\usepackage[normalem]{ulem}
\usepackage{comment}
\usepackage{xcolor}
\usepackage[caption=false]{subfig}
\usepackage{quantikz}

\newcommand*{\cnot}{\operatorname{CNOT}}
\newcommand*{\SWAP}{\operatorname{SWAP}}

\newcommand{\prob}[2]{\underset{#1}{\Pr}\left[#2\right]}

\usepackage{amsthm}
\usepackage{thm-restate}

\usepackage{ifdraft}
\ifdraft{
    \newcommand{\authnote}[3]{{\color{#3}[\textbf{#1:} #2]}}
}{
	\newcommand{\authnote}[3]{}
}

\begin{document}

\title{Repetition-code-based readout error detection and correction across hardware platforms and generations}

\author{Csaba Czab\'an}
\email{Contact author: czaban.csaba@wigner.hun-ren.hu}
\affiliation{HUN-REN Wigner Research Centre for Physics, H-1525   Budapest, Hungary}
\affiliation{HUN-REN Alfréd Rényi Institute of Mathematics, Budapest, Hungary}
\affiliation{ELTE Eötvös Loránd University, H-1117 Budapest, Hungary}
\author{Orsolya K\'alm\'an}%
\affiliation{HUN-REN Wigner Research Centre for Physics, H-1525   Budapest, Hungary}
\author{Sergey N. Filippov}
\affiliation{Algorithmiq Ltd., Helsinki, Finland}
\author{Zolt\'an Zimbor\'as}
\email{Contact author: zoltan.zimboras@helsinki.fi}
\affiliation{Department of Physics, University of Helsinki, Helsinki, Finland}
\affiliation{Algorithmiq Ltd., Helsinki, Finland}
\affiliation{HUN-REN Wigner Research Centre for Physics, H-1525   Budapest, Hungary}


\begin{abstract}
Readout errors are one of the dominant sources of noise in current quantum processors, limiting both expectation-value estimation and sampling-based applications. 
Since they affect only the classical measurement outcomes, they can be addressed using classical coding techniques: immediately before measurement, each data qubit is redundantly encoded with ancilla qubits, and the resulting bit string is decoded either by post-selection or by majority voting. Unlike conventional readout error mitigation, which corrects only aggregate quantities such as expectation values, this approach operates on individual measurement shots and can therefore produce approximately corrected samples. We present a systematic cross-platform and cross-generation experimental evaluation of repetition-code readout error detection and correction. We benchmark the same protocol on IBM Heron r1–r3 superconducting processors and Quantinuum H1 and H2 trapped-ion processors while independently varying the code distance, hardware generation, and encoding layout. We find that both error detection and correction improve readout fidelity on every device and generation tested, even as the unencoded baseline improves substantially across successive hardware releases. At the same time, the value of additional redundancy depends strongly on the underlying hardware. On superconducting processors, the extra gate errors introduced by the encoding rapidly offset its benefits, whereas on trapped-ion processors the much lower gate error rates allow larger code distances to remain advantageous.
\end{abstract}


\maketitle

\section{\label{sec:Intro}{Introduction}}

Noise is a central obstacle to extracting reliable results from present-day quantum processors. Of the relevant error channels, readout errors are among the most damaging. On current superconducting hardware, single-qubit assignment error rates are routinely on the order of one percent and often exceed two-qubit gate error rates. Although trapped-ion platforms typically achieve lower readout error rates, these remain higher than their two-qubit gate error rates.
Because every quantum computation terminates in measurement, readout errors act directly on the data ultimately returned to the user. Moreover, their impact grows with the number of measured qubits. As circuit widths increase, uncorrected readout errors therefore become a leading bottleneck for both expectation-value estimation and sampling-based applications.

A large body of work mitigates readout error through classical post-processing of the measured statistics. The standard strategy characterizes the imperfect measurement as a classical stochastic (assignment, or ``confusion'') matrix through a calibration procedure, and then inverts this matrix, or solves an associated constrained least-squares or maximum-likelihood problem, to recover corrected outcome probabilities or operator expectation values \cite{maciejewski2020mitigation, bravyi2021mitigating, chen2019detector,geller2020rigorous}. These methods are effective, and scalable variants avoid constructing the full assignment matrix explicitly \cite{nachman2020unfolding, Nation2021, maciejewski2021modeling, smith2021qubit,yang2022efficient,cosco2025bayesian,readout_arxiv_2026}. They share, however, a structural limitation that is central to the present work: they operate at the level of aggregate statistics rather than individual measurement records. The reconstructed object is an estimate of the full outcome distribution or of an expectation value; it is not a corrected set of shots. Moreover, the inverted distribution is in general only a quasi-distribution as it may contain negative entries, so it is not directly samplable without regularization, and reconstructing the full distribution is itself exponentially costly in the number of measured qubits. As a result, these techniques can return accurate expectation values, but they do not yield a stream of individual measurement outcomes distributed according to the ideal, noiseless measurement distribution. For any task that consumes individual samples rather than expectation values (such as sampling subroutines, distribution learning, or any procedure that post-selects on measured bitstrings) distribution-level mitigation is insufficient.

A complementary class of methods instead corrects readout error at the level of individual shots. Immediately before measurement, each data qubit is redundantly encoded by entangling it with one or more freshly prepared ancilla qubits through CNOT gates, mapping 
$\alpha_0\ket{0} + \alpha_1\ket{1} \mapsto \alpha_0\ket{0}^{\otimes n} + \alpha_1\ket{1}^{\otimes n}$
after which all qubits are measured in the computational basis. Because the encoding is applied only immediately before readout, it leaves the preceding computation untouched and targets precisely the computational-basis bit-flip errors that dominate measurement noise. Two strategies then follow. In readout error \emph{detection}, one post-selects on the data and ancilla outcomes agreeing (a unanimous vote) and discards any record in which they disagree, since such a disagreement signals that a readout error has occurred. In readout error \emph{correction}, one retains every record and assigns the logical outcome by majority vote over the root (data) qubit and its ancillas. Both strategies act shot by shot and therefore return individually corrected measurement records, and hence approximately correct samples, at the cost of a reduced number of retained shots or of residual mis-corrections, respectively. This repetition-code approach to readout, in both its detection and correction forms, has been introduced and analyzed in prior works \cite{readout_QuantSciTech_2021,readout_PhysRevA_2022,readout_PhysRevRes_2023,ouyang2024robust, linden2025use,byrne2026reducing}. However, these studies have been either entirely theoretical or confined to individual superconducting devices, leaving open how the scheme performs across qualitatively different hardware. In particular, it remains unexplored how the method performs on trapped-ion processors, whose error characteristics differ substantially from those of superconducting platforms, or how its benefit evolves across successive hardware generations. 

In this work, we address these questions by
by evaluating repetition-code readout error detection and correction on IBM Heron r1, r2, and r3 devices and Quantinuum H1 and H2 trapped-ion quantum processor units. After briefly recalling the basics of these methods in Sec.~\ref{sec:voting-meas}, we present our analysis on their experimental implementation in Sec.~\ref{sec:experiments}. We summarize our main conclusions in Sec.~\ref{sec:Concl}.

\section{\label{sec:voting-meas}{Readout error detection and correction with repetition codes}}

\subsection{Repetition code and decoding rules}
\label{sec:repcode-rules}

The binary repetition code is the simplest classical linear code. It encodes a
single logical bit into $n$ physical bits by repetition,
\begin{equation}
  0 \longmapsto \underbrace{00\cdots 0}_{n}, \qquad
  1 \longmapsto \underbrace{11\cdots 1}_{n},
  \label{eq:codewords}
\end{equation}
so that the codebook consists of the two constant strings,
$\mathcal{C}=\{0^{n},1^{n}\}$. Regarded as a linear code over $\mathrm{GF}(2)$,
the $n$-fold repetition code is, in the standard $[n,k,d]$ notation for a code
of length $n$, dimension $k$, and minimum Hamming distance $d$, the $[n,1,n]$
code, i.e., it has length $n$, encodes $k=1$ logical bit (two codewords), and has
minimum distance $d=n$, since the two codewords differ in every position. It is
in fact the unique binary linear code with these parameters, as any $[n,1]$
linear code is generated by a single nonzero codeword whose weight equals the
minimum distance. Thus, we adopt the
usual shorthand $[n,1]$ for the $n$-repetition code.

A code of minimum distance $d$ can detect up to $d-1$ errors, or correct up to
$\lfloor (d-1)/2 \rfloor$ errors. For the $[n,1]$ code these bounds correspond to $n-1$
detectable and $\lfloor (n-1)/2 \rfloor$ correctable bit flips. We realize the two operating modes through two distinct decoding rules applied to a received $n$-bit word 
$\mathbf{b}=(b_{1},\dots,b_{n})$.

\emph{Unanimous-vote decoding.} The word is accepted as error-free only if all
of its bits agree, i.e., if $\mathbf{b}\in\{0^{n},1^{n}\}$; any disagreement flags a
detected error. This rule attains the full detection distance: it flags every
error pattern of weight between $1$ and $n-1$, failing only in the extreme case
in which all $n$ bits are flipped and the received word coincides with the
opposite codeword. A flagged word carries no reliable logical value, so the
natural response is to discard it. For a memoryless bit-flip channel with
crossover probability $p$ acting independently on each position, an accepted
word is mis-assigned only if every bit is flipped onto the opposite codeword,
so the residual error among accepted words scales as follows for small $p$
\begin{equation*}
    \prob{}{\text{error}\mid\text{accept}} = \frac{p^n}{(1-p)^n + p^n} \;\sim\; \left(\frac{p}{1-p}\right)^n \sim p^n,
\end{equation*}
i.e., decreases exponentially with $n$. The price is that a growing fraction of words is rejected.

\emph{Majority-vote decoding.} The logical bit is assigned the value held by the
majority of the $n$ positions. For odd $n$ this is always well defined and
corrects any error of weight up to $(n-1)/2$, whereas for even $n$ a tie can
occur on exactly $n/2$ flips. We therefore take $n$ odd whenever the code is
used for correction. Unlike unanimous-vote decoding, majority vote returns a
logical value for every received word and discards nothing. Under the same
independent bit-flip channel, the per-word logical error requires the majority
of the bits to flip, i.e., at least $(n+1)/2$ of the $n$ bits. The logical error
is thus given by
\begin{equation*}
   \prob{}{\text{error}} = \sum_{w=(n+1)/2}^{n} \binom{n}{w}\, p^{w}(1-p)^{n-w},
\end{equation*}
which at small $p$ is dominated by it's lowest weight-term,
\begin{equation*}
\binom{n}{(n+1)/2}p^{\,(n+1)/2} \sim p^{\,(n+1)/2},
\end{equation*}
a slower suppression than for detection but achieved without
rejecting any words.

The two rules thus sit at opposite ends of the detection--correction trade-off
intrinsic to any code: unanimous-vote decoding maximizes the number of errors
flagged but recovers nothing from a flagged word, while majority-vote decoding
always returns a value at the cost of tolerating fewer errors before
mis-decoding.

\subsection{Application to readout error detection and correction}
\label{sec:repcode-readout}

The code~\eqref{eq:codewords} is classical, and 
it cannot serve as a general-purpose quantum error-correcting code. Applied to a
qubit, the encoding maps
\begin{equation}
  \alpha_{0}\ket{0}+\alpha_{1}\ket{1}
  \;\longmapsto\;
  \alpha_{0}\ket{0}^{\otimes n}+\alpha_{1}\ket{1}^{\otimes n},
  \label{eq:encoding}
\end{equation}
which is realized by $\mathrm{CNOT}$ gates that copy the data qubit's
computational-basis value onto $n-1$ freshly initialized ancillas. This is not
cloning: it copies the value only in the computational basis, producing an
entangled (GHZ-type) state rather than $n$ independent copies, without violating
the no-cloning theorem. As a quantum code, \eqref{eq:encoding} is the bit-flip
repetition code, which protects only against $X$-type errors; a single phase
($Z$) error on any qubit is neither detected nor corrected, and protecting an
arbitrary qubit against general errors requires a genuine quantum code such as
the $[[5,1,3]]$ five-qubit Laflamme  code or the $[[9,1,3]]$ Shor  code~\cite{albert2026handbookerrorcorrectingcodes}. We therefore do not use
\eqref{eq:encoding} for quantum error correction.

Its use here is narrower. We apply the encoding immediately before a
computational-basis measurement, and the quantity we wish to protect is not an
arbitrary quantum state but the classical computational-basis label that the
measurement is meant to return. Readout error is, to a
good approximation, a classical bit-flip acting independently on each measured
outcome. Under the encoding of Eq.~\ref{eq:encoding}, the single logical value is spread across $n$ qubits that
are then measured independently, yielding $n$ noisy classical copies of that
value, which is precisely the setting of the classical $[n,1]$ repetition code. The
decoding rules of Sec.~\ref{sec:repcode-rules} apply directly to the $n$
measured bits: unanimous-vote decoding detects readout errors and post-selects,
and majority-vote decoding corrects them. The phase errors the code cannot
handle are irrelevant here, since a $Z$ error immediately before a
computational-basis measurement does not change the measured bit. This is why a
code inadequate for quantum error correction is nonetheless well matched to
readout: the readout channel is essentially a classical bit-flip channel, which
is exactly what the repetition code addresses.

The idealized picture of $n$ independent noisy copies of one bit is exact only
in the absence of encoding faults and correlated readout errors. In practice the
$n-1$ encoding $\mathrm{CNOT}$s are themselves imperfect, injecting errors before
the measurement, and these can be correlated across the encoded qubits. Whether
the redundancy supplied by~\eqref{eq:encoding} still reduces the readout error
once this cost is included is the central empirical question of this work. An additional consideration is the choice of encoding layout: the 
$\mathrm{CNOT}$s can be arranged in different layouts that prepare the
same encoded state but differ in their connectivity requirements and in how
encoding faults propagate; we defer the description of these layouts to
Sec.~\ref{sec:experiments}\,A, where additional topology constraints are also taken into account.

\section{\label{sec:experiments}{Experimental results}}


In this section, we detail the experiments we conducted concerning the detection and correction of readout errors described in \Cref{sec:voting-meas}. The experiments were carried out on IBM's superconducting devices, spanning the Heron processor family across generations r1 to r3 and on Quantinuum's trapped-ion \mbox{H-series} (H1-1 and H2-1). We first describe how the repetition-code encodings are implemented on each platform in \Cref{sec:gen}. We then evaluate our findings along several axes. We begin by analyzing the error-detection results from unanimous-vote decoding, i.e.\ post-selecting on the shots in which the data qubit and all of its ancillas agree. In \Cref{sec:code_dist} we examine how increasing the code distance reshapes the cumulative distribution of post-selected readout errors; in \Cref{sec:gens} we compare the readout error with and without post-selection across successive device generations; and in \Cref{sec:code_tops} we compare the two encoding topologies, star versus chain, for the 
[3,1] encoding. Finally, in \Cref{sec:err_corr} we turn from detection to correction: rather than discarding faulty shots, we recover them by majority-vote decoding, and compare the readout error after correction against the unmitigated baseline.

\subsection{Implementing repetition-code encodings on the experimental platforms}
\label{sec:gen}  

\begin{figure}[th]
  \centering
  \begin{minipage}[b]{0.49\linewidth}
    \centering
    \resizebox{\linewidth}{!}{%
    \begin{quantikz}[row sep={0.6cm,between origins}, column sep=0.34cm]
      \lstick{$q_{\mathrm{data}}\;\ket{\psi}$} & \ctrl{1} & \ctrl{2} & \ctrl{3} & \meter{} \\
      \lstick{$a_1\;\ket{0}$} & \targ{} &         &         & \meter{} \\
      \lstick{$a_2\;\ket{0}$} &         & \targ{} &         & \meter{} \\
      \lstick{$a_3\;\ket{0}$} &         &         & \targ{} & \meter{} \\
    \end{quantikz}}
    
    \vspace{0.2cm}
    (a) star layout  
  \end{minipage}\hfill
  \begin{minipage}[b]{0.49\linewidth}
    \centering
    \resizebox{\linewidth}{!}{%
    \begin{quantikz}[row sep={0.6cm,between origins}, column sep=0.34cm]
      \lstick{$q_{\mathrm{data}}\;\ket{\psi}$} & \ctrl{1} &         &         & \meter{} \\
      \lstick{$a_1\;\ket{0}$} & \targ{} & \ctrl{1} &         & \meter{} \\
      \lstick{$a_2\;\ket{0}$} &         & \targ{} & \ctrl{1} & \meter{} \\
      \lstick{$a_3\;\ket{0}$} &         &         & \targ{} & \meter{} \\
    \end{quantikz}}
    
    \vspace{0.2cm}
    (b) chain layout 
  \end{minipage}
  
  \caption{
  Circuit diagrams for implmenting the $[3,1]$ repetition code with two CNOT gates using (a) star layout and (b) chain layout. 
  }
  \label{fig:circuits}
\end{figure}
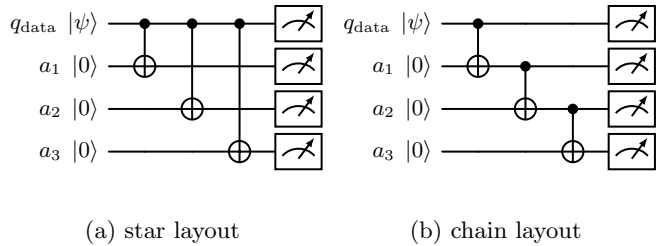

To realize the $[n,1]$ repetition-code encoding of~\Cref{eq:encoding}, we use $n{-}1$ ancilla qubits and $n{-}1$ $\cnot$ gates, arranged in one of two distinct layouts that differ in how the data qubit connects to the ancillas. In the \emph{star} layout, each  $\cnot$ is applied from the data qubit to an individual ancilla. In the \emph{chain} layout, the $\cnot$s are applied sequentially: first from the data qubit to the first ancilla, then from the first ancilla to the second, and so on. The two layouts are shown in~\Cref{fig:circuits}.

\begin{figure}[tb]
\includegraphics[width=\linewidth]{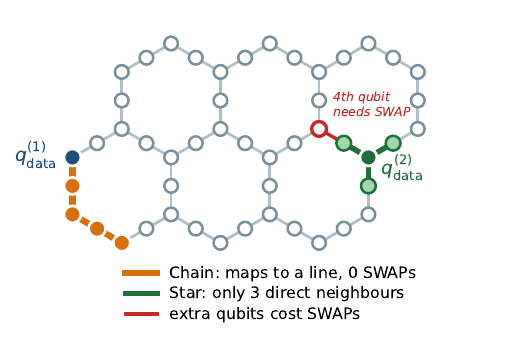}
  \caption{ The implementation of the chain (orange) and star (green) layouts in the heavy-hex topology. Note that the star layout is limited to at most three ancillas, as the addition of a fourth one would require extra SWAP gates (red).}
  \label{fig:heavy-hex-topology}
\end{figure}

While we implement both layouts across both platforms, device connectivity imposes significant practical constraints. Quantinuum's trapped-ion processors offer all-to-all connectivity, allowing the \emph{star} layout to scale without routing overhead. However, on IBM's heavy-hexagonal lattice, a given qubit is physically connected to a maximum of three neighbors. Consequently, implementing the \emph{star} layout on these superconducting devices restricts us to at most three ancillas since scaling further requires inserting $\SWAP$ gates. The \emph{chain} layout avoids this problem entirely, mapping naturally onto the linear segments of the heavy-hex lattice without routing overhead, making it a highly advantageous topology for devices lacking all-to-all connectivity as shown in~\Cref{fig:heavy-hex-topology}.

In each experiment, we prepare the data qubits in a computational basis state, either all-zeros, all-ones, or a random bitstring, and read them out. For these inputs, we evaluate the per-qubit (marginal) readout error, computed qubit by qubit, for three cases: without any encoding; with repetition-code encoding decoded by unanimous vote, where shots in which a data qubit and its ancillas disagree are post-selected away (error detection); and with repetition-code encoding decoded by majority vote, where every shot is retained and faulty outcomes are corrected rather than discarded (error correction). Because post-selection discards shots, we additionally investigate the fraction of samples rejected under unanimous-vote decoding when the full bitstring is read out, i.e., when a shot is kept only if every data qubit passes its post-selection simultaneously, which quantifies the sampling overhead of the detection scheme for whole-register readout.

\subsection{Effect of code distance on post-selected readout errors}
\label{sec:code_dist}


We first investigate how the code distance affects the readout errors. Consider an idealized situation in which the encoding $\cnot$s are noiseless. In this case, increasing the code distance can only help: a larger $n$ in the $[n,1]$ repetition code spreads the logical value across more physical qubits, so that under unanimous-vote decoding a larger fraction of readout errors produces a detectable disagreement and is post-selected away. As discussed in \Cref{sec:voting-meas}, for independent readout errors, the residual error among the accepted shots therefore decreases exponentially with $n$, and one would expect the post-selected readout error to be driven toward zero as the code distance grows.

\begin{figure}[t!]
    \centering
    \includegraphics[width=\linewidth]{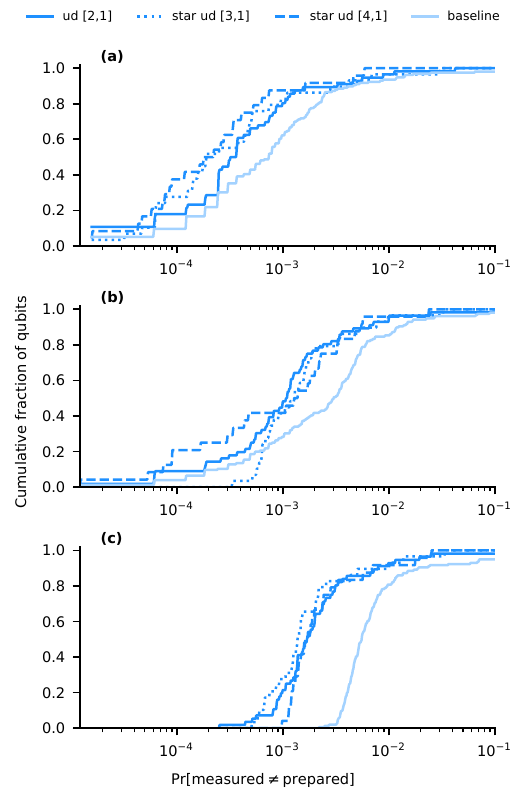}
    \caption{The cumulative fraction of qubits versus the estimated probability of detecting an error per qubit for the case of  the $[2,1]$, $[3,1]$, and $[4,1]$ codes decoded by unanimity in the star layout on IBM Pittsburgh for three different input initializations (a) all-zeros, (b) random, and (c) all-ones.
    }
    \label{fig:code-distance-pittsburgh}
\end{figure}

This conclusion is fragile, however, because the encoding $\cnot$s are not noiseless. Each unit increase in the code distance adds one ancilla qubit and one $\cnot$ gate, injecting additional two-qubit-gate error. These extra errors act in the opposite direction: they cause genuine outcomes to be rejected, and more problematically, they can produce correlated faults that are accepted as valid codewords, contributing to the residual error rather than being filtered out. Whether increasing $n$ actually reduces the post-selected readout error in practice is therefore not clear a priori. It depends on the balance between the readout error, which the redundancy suppresses, and the two-qubit-gate error, which grows with $n$. This trade-off is precisely what the following experiments probe. To this end, we report the post-selected readout error as a function of code distance on two representative platforms with markedly different error profiles: for the superconducting case we use IBM Pittsburgh, a Heron r3 device, while for the trapped-ion case we use Quantinuum H1-1. 

\begin{figure}[th]
    \centering
    \includegraphics[width=\linewidth]{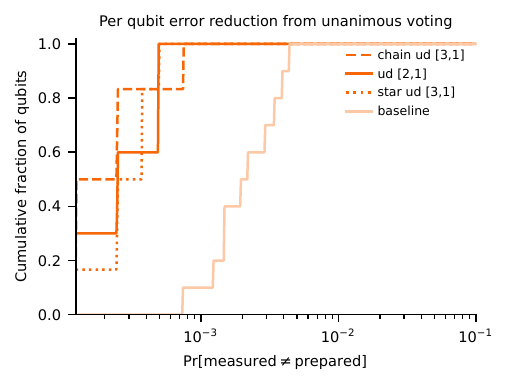}
    \caption{The cumulative fraction of qubits versus the estimated probability of detecting an error per qubit for the case of the $[2,1]$ and $[3,1]$ encodings in the star and chain layouts, with unanimous-vote decoding (ud) on Quantinuum's H-series devices for a random initialization of the qubits.}
    \label{fig:code-distance-h1}
\end{figure}

\begin{figure*}[tb]
    \renewcommand\thesubfigure{(a--c)}
    \subfloat[\label{fig:acrros-gen_IBM}]{        \includegraphics[width=0.98\linewidth]{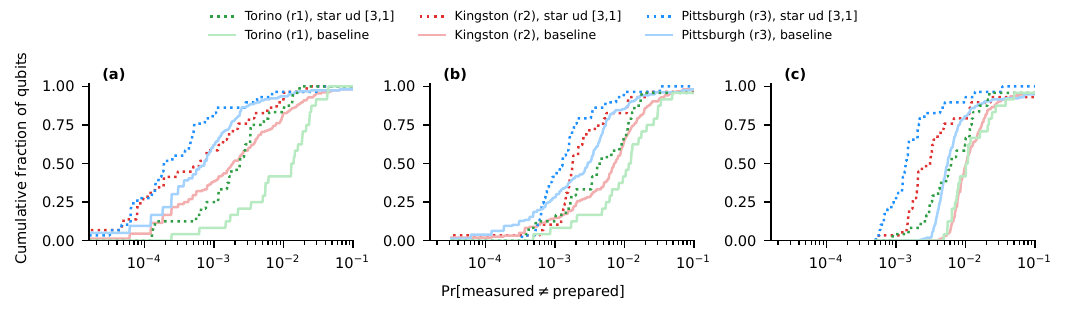}
    } \\
    \vspace{-5ex}
    \renewcommand\thesubfigure{(d--f)}
    \subfloat[\label{fig:acrros-gen_Quantinuum}]{        \includegraphics[width=0.98\linewidth]{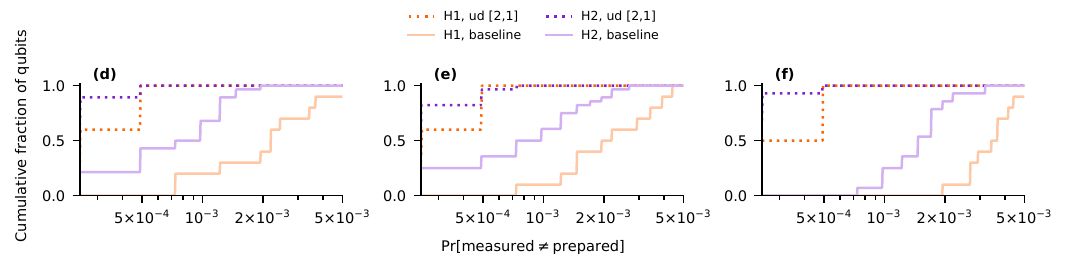}
    }
    \vspace{-4ex}
    \caption{The cumulative fraction of qubits versus the estimated probability of detecting an error per qubit. Panels (a--c) use the star layout and the unanimous-vote decoding (ud) in the $[3,1]$ code for three generations of IBM Heron processors: r1 (Torino), r2 (Kingston), and r3 (Pittsburgh). Panels (d--f) show results for Quantinuum's H1-1 and H2-1 trapped-ion devices using the same decoding in the $[2,1]$ code. Subfigures correspond to three different input initializations: (a) and (d) all-zeros, (b) and (e) random, and (c) and (f) all-ones. Baselines are shown for reference.
    }
    \label{fig:acrros-gen}
\end{figure*}

\Cref{fig:code-distance-pittsburgh} reports results for IBM Pittsburgh, a Heron r3 device. Its three panels show the empirical cumulative distribution of the per-qubit readout error across the qubits of the device, for the three input states (all-zeros, all-ones, and a random computational-basis string), comparing the unencoded baseline against the $[2,1]$, $[3,1]$, and 
$[4,1]$ codes decoded by unanimous vote, all prepared in the star layout. For every input and every code distance, error detection shifts the distribution toward lower error rates relative to the baseline, so that the readout error of a typical qubit is reduced by post-selection. The dependence on code distance, however, varies markedly with the input. For the all-zeros input the distributions show a slight improvement monotonically with distance, in line with the idealized expectation. For the all-ones and random inputs, by contrast, the three code distances perform almost identically: the benefit of error detection is essentially saturated already at 
$[2,1]$, and adding further ancillas yields no appreciable improvement. The gain from added redundancy appears to be cancelled by the extra gate error, so increasing the code distance brings no further improvement for these inputs on this device.

\Cref{fig:code-distance-h1} shows the corresponding results for the Quantinuum H1-1 trapped-ion device, for a random computational-basis input. It compares the cumulative distribution of the readout error for the unencoded baseline against those obtained after unanimous-vote decoding of the 
$[2,1]$ and $[3,1]$ codes, where for the latter we include both the star and the chain layout. Here the ordering is cleaner and matches the idealized picture: the $[2,1]$ code already surpasses the baseline, and the 
$[3,1]$ code typically reduces the readout error further still, so that on this platform increasing the code distance continues to help rather than saturating. The contrast with the random-input behavior on Pittsburgh is the central observation of this section: the same code distance buys a further improvement on the trapped-ion device but not on the superconducting one. This is consistent with the trade-off set out above: Quantinuum's substantially lower two-qubit-gate error relative to its readout error keeps the device in the regime where redundancy is still winning, whereas on the superconducting device the gate-error cost of each additional ancilla has already caught up with the benefit. (The two 
$[3,1]$ layouts are shown here for completeness; we compare star and chain in detail in \Cref{sec:code_tops}.)

\subsection{Effectiveness of readout error detection across device generations}
\label{sec:gens}

Within a given hardware family, successive generations improve essentially all error parameters at once: newer devices typically exhibit both lower readout error and lower two-qubit-gate error. For the repetition-code scheme these two improvements pull in opposite directions. Lower readout error reduces the very quantity the redundancy is designed to suppress, shrinking the room for error detection to help; lower gate error reduces the cost each added ancilla incurs, making the redundancy cheaper to deploy. The effectiveness of error detection on a newer device therefore depends not on either quantity in isolation but on how the readout error and the gate error move relative to each other from one generation to the next. If readout error falls faster than gate error, the scheme has less to correct and its benefit may erode; if gate error falls faster, the favorable regime widens and the benefit may grow. Which of these occurs is an empirical question about the trajectory of each hardware family.

To probe this, we benchmark error detection across the generations of two platforms, comparing the unencoded baseline against unanimous-vote decoding on each device. For the IBM superconducting family we use the three Heron generations (r1, r2, and r3) with the  $[3,1]$ code. For the Quantinuum trapped-ion family we use the H1-1 and H2-1 devices with the $[2,1]$ code. The two platforms are analyzed as separate within-family progressions rather than on a common axis, since they differ both in code distance and in underlying technology; the comparison of interest is how the benefit of error detection evolves along each family's generational sequence.

\Cref{fig:acrros-gen} summarizes the generational comparison.  The upper row shows the resulst for the $[3,1]$ code in the star layout across three Heron generations: r1 (Torino), r2 (Kingston), and r3 (Pittsburgh). The lower row uses the $[2,1]$ code across the Quantinuum H1-1 and H2-1 generations.

Two trends are visible simultaneously, and their coexistence is the main point of the figure. First, the baseline curves shift toward lower error with each successive generation: on both platforms the unencoded readout error of a typical qubit improves from the older to the newer device, confirming the expected hardware progression. Second, and despite this moving baseline, error detection continues to deliver a clear improvement on every generation: in each panel the post-selected curve lies below the corresponding baseline, so the scheme produces better readout statistics than the bare measurement even on the most recent devices, where the baseline is already at its best.

\subsection{Star versus chain: effect of layout on error detection}
\label{sec:code_tops}

\begin{figure}[t]
    \centering
    \includegraphics[width=\linewidth]{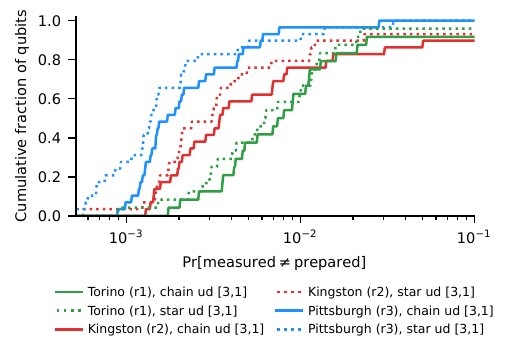}
    \caption{Comparison of error detection efficiency with star and chain layouts  on different generations of IBM Heron processors, with qubits initialized in the all-ones state. Coloring is the same as in Fig.~\ref{fig:acrros-gen_IBM}.}
    \label{fig:star-vs-chain-ibm-uv}
\end{figure}

\begin{figure}[t]
    \centering
    \includegraphics[width=\linewidth]{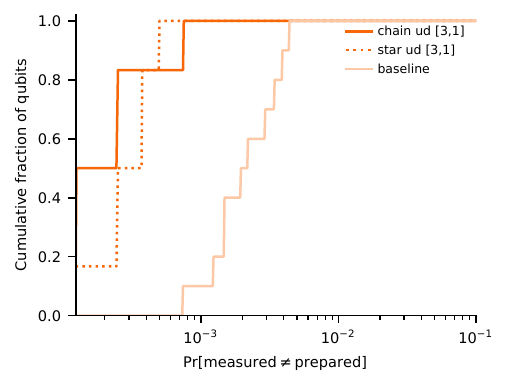}
    \caption{Comparison of error detection efficiency with star and chain layouts  on different generations of Quantinuum processors, with qubits initialized in the all-ones state. Coloring is the same as in Fig.~\ref{fig:acrros-gen_Quantinuum}.}
    \label{fig:star-vs-chain-quantinuum-uv}
\end{figure}

We now isolate the effect of the encoding layout, comparing the star and chain arrangements at fixed code distance 
$[3,1]$. To stress the scheme, we use the all-ones input throughout this comparison as in the superconducting devices the readout error is larger with such inputs.  

\Cref{fig:star-vs-chain-ibm-uv} reports the three Heron generations r1--r3 and \Cref{fig:star-vs-chain-quantinuum-uv} the Quantinuum H1-1 device. The two platforms behave differently. On all three Heron generations the star layout outperforms the chain, producing lower post-selected readout error. On the Quantinuum H1-1 device, by contrast, the two layouts perform comparably, with no marked advantage for either. The layout choice thus matters on the superconducting hardware but not on the trapped-ion device.

\subsection{From error detection to overhead-free majority-vote correction}\label{sec:err_corr}

The error-detection scheme examined so far improves readout at a cost: every shot in which the data qubit and its ancillas disagree is discarded, so the gain in accuracy is paid for in retained statistics. Under unanimous-vote decoding this cost compounds across the register, since a shot is kept only if every encoded qubit passes its check simultaneously. As a result, the fraction of shots that survive post-selection falls off exponentially with the number of qubits read out: each additional qubit multiplies the surviving fraction by its own acceptance probability, so even modest per-qubit rejection rates accumulate into a severe overhead for wide registers. For tasks that consume individual samples, this sampling overhead quickly outweighs the benefit of cleaner readout. Majority-vote decoding offers a way out. For the $[n,1]$ repetition code with $n$ odd, every measured outcome can be assigned a logical value by majority over the  $n$ readout values, so no shot is ever ambiguous and none is discarded. Correction in this sense is overhead-free: it retains the full shot budget while still suppressing readout errors,  at the cost of a higher residual error than post-selection, since ambiguous outcomes are corrected rather than removed.

\begin{table}[h!]
  \centering
  \begin{tabular}{llcccc}
    \toprule
    Processor & Layout & Ancillas & \makecell{Median\\retention} & \makecell{Mean\\retention} & \makecell{Joint\\retention} \\
    \midrule
    Pittsburgh & star & 1 & 98\% & 98\% & 55\% \\
    Pittsburgh & chain & 2 & 97\% & 95\% & 29\% \\
    Pittsburgh & star & 2 & 96\% & 96\% & 34\% \\
    Pittsburgh & star & 3 & 95\% & 91\% & 4\% \\
    \midrule
    H1 & star & 1 & 99\% & 99\% & 96\% \\
    H1 & chain & 2 & 99\% & 99\% & 94\% \\
    H1 & star & 2 & 99\% & 99\% & 94\% \\
    \midrule
  \end{tabular}
  \caption{Unanimous-voting postselection retention on IBM's Pittsburgh (r3) and Quantinuum's H1-1 machines for the random preparations. To compare runs of different widths fairly within the same machine, every run is cut to the smallest data-qubit count by drawing that many data qubits at random without replacement, and all statistics are computed over that common sub-block.}
\label{tab:across-uv-retention}
\end{table}

\begin{figure}[b]
    \centering
    \includegraphics[width=\linewidth]{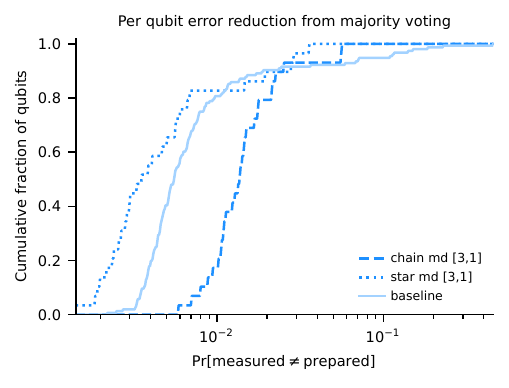}
    \caption{The cumulative fraction of qubits versus the estimated probability of correcting an error per qubit via majority decoding, for the star and chain layouts with the qubits initialized in the all-ones computational basis state on IBM's Pittsburgh processor.}
    \label{fig:pittsburgh-random}
\end{figure}

\begin{figure}[b]
    \centering
    \includegraphics[width=\linewidth]{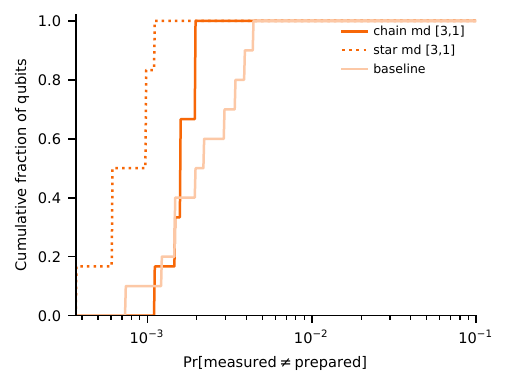}
    \caption{The cumulative fraction of qubits versus the estimated probability of correcting an error per qubit via majority decoding, for the star and chain layouts with each qubit initialized to a random quantum state on Quantinuum's H1-1 processor.}
    \label{fig:h1-random}
\end{figure}

We first quantify this overhead directly. \Cref{tab:across-uv-retention} reports the fraction of shots retained under unanimous-vote decoding on IBM Pittsburgh, using $24$ data qubits, for the $[2,1]$, $[3,1]$, and 
$[4,1]$ codes. The retained fraction falls sharply with code distance: the 
$[3,1]$ code keeps only about a third of the shots for both the star and chain layouts, and the $[4,1]$ code keeps roughly $4\% $. In other words, recovering one clean sample from the 
$[4,1]$-encoded register requires on the order of twenty-five raw shots, a substantial penalty that grows with both the code distance and the number of qubits read out. The lower part of \Cref{tab:across-uv-retention} reports the corresponding figures for the Quantinuum H1-1 device with 6 data qubits, for the $[2,1]$ and $[3,1]$ cases. Here the picture is far more favorable: over $90\%$ of the shots are retained. The contrast reflects the two factors: H1-1 has lower readout error per qubit and and we could only use a narrower register, so far fewer shots fail the unanimity check.

We now turn to majority-vote correction, which incurs no shot overhead, and ask whether it still improves on the unencoded readout. We compare the star and chain layouts for the $[3,1]$ code with the qubits initialized to the all-ones computational-basis state, on IBM Pittsburgh  and Quantinuum H1-1, against the unencoded baseline. \Cref{fig:pittsburgh-random} shows, for Pittsburgh, the cumulative fraction of qubits as a function of the estimated per-qubit error probability after majority-vote correction, for both layouts. The two layouts behave qualitatively differently. The star layout improves on the baseline: correcting by majority over the data qubit and its two ancillas yields a lower per-qubit error than the bare measurement. The chain layout, however, performs worse than the baseline—majority-vote correction in this arrangement increases the per-qubit error rather than reducing it.

This is the central caution of the section. Unlike post-selection, which discards faulty shots and can therefore in principle only improve the accepted readout, majority-vote correction keeps every shot and commits to a decoded value; when the encoding is too noisy, that commitment can do net harm. In the chain layout the two 
$\cnot$s are applied in sequence, so a fault on the first ancilla propagates to the second, producing correlated errors that flip the majority outcome rather than being outvoted. On a device where the encoding gate error is non-negligible, this suffices to push the chain-corrected readout below the baseline, whereas the star layout, with both 
$\cnot$s controlled by the data qubit, does not chain faults in the same way and remains a net improvement.

\Cref{fig:h1-random} shows the corresponding results for Quantinuum H1-1. Here both layouts improve on the baseline: the trapped-ion device's substantially lower two-qubit-gate error means that even the chain arrangement introduces few enough correlated faults that the majority vote remains beneficial. The contrast with Pittsburgh underscores that majority-vote correction is only advantageous when the encoding gate error is small relative to the readout error it is meant to suppress, a condition the trapped-ion device satisfies comfortably for both layouts, but which the superconducting device satisfies only for the star.

\section{\label{sec:Concl}Conclusion and Outlook}

We have presented a systematic experimental evaluation of repetition-code–based readout error detection and correction across two qubit technologies and several generations of hardware. Specifically, we carried out controlled comparisons of the same scheme on IBM Heron r1–r3 superconducting processors and Quantinuum H1 and H2 trapped-ion processors, varying code distance, device generation, and encoding layout as independent parameters. In addition, we directly measured the overhead incurred by post-selection.

Four observations emerged. First, the benefit of increasing the code distance is strongly device dependent. On the superconducting device the post-selected error saturates already at the $[2,1]$ code for the all-ones and random inputs, the gain from further redundancy being offset by the gate error of the added encoding $\mathrm{CNOT}$s, whereas on the trapped-ion device, where the two-qubit-gate error is far smaller relative to the readout error, increasing the distance continues to help. Second, across the generations examined, error detection remained beneficial on every device: although the unencoded baseline improved from one generation to the next on both platforms, post-selection still produced a net reduction in readout error even on the most recent hardware. Third, the encoding layout matters on the superconducting devices, where the star arrangement outperformed the chain on all three Heron generations, but not on the trapped-ion device, where the two layouts performed comparably. Fourth, the two decoding strategies sit at opposite ends of a practical trade-off: unanimous-vote detection lowers the readout error but discards shots at a rate that grows sharply with both code distance and register width, retaining only a few percent of whole-register samples for the $[4,1]$ code on $24$ qubits, while majority-vote correction discards nothing but, being committed to a decoded value, can do net harm when the encoding is too noisy, as seen for the chain layout on the superconducting device (IBM Pittsburgh), where it pushed the corrected readout below the unencoded baseline. 

The implication of these findings is that the scheme is not uniformly advantageous: its configuration (code distance, layout, and the choice between detection and correction) should be matched to the measured error profile of the specific device rather than applied by default.

Several caveats point to future work. Our per-qubit metric does not resolve correlated readout errors, and the mechanism we propose for the chain layout's failure under correction is inferred rather than measured; both warrant direct
characterization, as does whether the benefit of detection erodes as readout error falls over a longer generation sequence of devices. 
One of the most promising directions opened by this work is the development of adaptive implementations, in which the encoding is tailored to calibration data. In such an approach, encoding would be applied selectively, only to qubits for which it provides a net benefit, while error correction, rather than error detection, would be used only when the associated gate error is sufficiently low to make correction advantageous. Such an adaptive strategy would also help control the post-selection overhead, which remains the principal limitation of detection-based methods for sampling tasks on wide registers. Natural extensions include asymmetry-aware decoding, in which the vote is weighted according to the measured readout-error asymmetries.


\begin{acknowledgments}
This work was supported by the Horizon Europe programme HORIZON-CL4-2022-QUANTUM-01-SGA via the project 101113946 OpenSuperQPlus100, and by the European Union’s Horizon Europe research and innovation program under grant agreement No 101135699 via the project SPINUS. We also acknowledge support by the National Research, Development and Innovation Office through the AI4QT project Nr. 2020-1.2.3-EUREKA-2022-00029. 
C.C. was supported by the Lend\"{u}let ``Momentum'' program of the Hungarian Academy of Sciences under grant agreement no. LP2025-8/2025.
\end{acknowledgments}



\bibliography{readout}

@PREAMBLE{
 "\providecommand{\noopsort}[1]{}" 
 # "\providecommand{\singleletter}[1]{#1}%" 
}

@article{readout_PhysRevRes_2023,
  title = {Single-shot error mitigation by coherent Pauli checks},
  author = {van den Berg, Ewout and Bravyi, Sergey and Gambetta, Jay M. and Jurcevic, Petar and Maslov, Dmitri and Temme, Kristan},
  journal = {Phys. Rev. Res.},
  volume = {5},
  issue = {3},
  pages = {033193},
  numpages = {24},
  year = {2023},
  month = {Sep},
  publisher = {American Physical Society},
  doi = {10.1103/PhysRevResearch.5.033193},
  url = {https://link.aps.org/doi/10.1103/PhysRevResearch.5.033193}
}

@article{readout_PhysRevA_2022,
  title = {Active readout-error mitigation},
  author = {Hicks, Rebecca and Kobrin, Bryce and Bauer, Christian W. and Nachman, Benjamin},
  journal = {Phys. Rev. A},
  volume = {105},
  issue = {1},
  pages = {012419},
  numpages = {13},
  year = {2022},
  month = {Jan},
  publisher = {American Physical Society},
  doi = {10.1103/PhysRevA.105.012419},
  url = {https://link.aps.org/doi/10.1103/PhysRevA.105.012419}
}

@article{readout_QuantSciTech_2021,
doi = {10.1088/2058-9565/ac3386},
url = {https://doi.org/10.1088/2058-9565/ac3386},
year = {2021},
month = {nov},
publisher = {IOP Publishing},
volume = {7},
number = {1},
pages = {015009},
author = {G\"{u}nther, Jakob M and Tacchino, Francesco and Wootton, James R and Tavernelli, Ivano and Barkoutsos, Panagiotis Kl},
title = {Improving readout in quantum simulations with repetition codes},
journal = {Quantum Science and Technology}
}

@misc{albert2026handbookerrorcorrectingcodes,
      title={Handbook of Error-Correcting Codes}, 
      author={Victor V. Albert and Philippe Faist},
      year={2026},
      eprint={2606.11484},
      archivePrefix={arXiv},
      primaryClass={quant-ph},
      url={https://arxiv.org/abs/2606.11484}, 
}

@article{chen2019detector,
  title={Detector tomography on IBM quantum computers and mitigation of an imperfect measurement},
  author={Chen, Yanzhu and Farahzad, Maziar and Yoo, Shinjae and Wei, Tzu-Chieh},
  url={http://dx.doi.org/10.1103/PhysRevA.100.052315},
  DOI={10.1103/physreva.100.052315},
  journal={Physical Review A},
  volume={100},
  number={5},
  pages={052315},
  year={2019},
  publisher={APS}
}

@article{geller2020rigorous,
  title={Rigorous measurement error correction},
  author={Geller, Michael R},
  url={http://dx.doi.org/10.1088/2058-9565/ab9591},
  DOI={10.1088/2058-9565/ab9591},
  journal={Quantum Science \& Technology},
  volume={5},
  number={3},
  pages={03LT01},
  year={2020},
  publisher={IOP Publishing}
}

@article{nachman2020unfolding,
  title={Unfolding quantum computer readout noise},
  author={Nachman, Benjamin and Urbanek, Miroslav and de Jong, Wibe A and Bauer, Christian W},
  url={https://doi.org/10.1038/s41534-020-00309-7},
  DOI={10.1038/s41534-020-00309-7},
  journal={npj Quantum Information},
  volume={6},
  number={1},
  pages={84},
  year={2020},
  publisher={Nature Publishing Group UK London}
}

@article{maciejewski2021modeling,
  title={Modeling and mitigation of cross-talk effects in readout noise with applications to the Quantum Approximate Optimization Algorithm},
  url={http://dx.doi.org/10.22331/q-2021-06-01-464},
  DOI={10.22331/q-2021-06-01-464},
  author={Maciejewski, Filip B and Baccari, Flavio and Zimbor{\'a}s, Zolt{\'a}n and Oszmaniec, Micha{\l}},
  journal={Quantum},
  volume={5},
  pages={464},
  year={2021}
}

@article{yang2022efficient,
  title={Efficient quantum readout-error mitigation for sparse measurement outcomes of near-term quantum devices},
  author={Yang, Bo and Raymond, Rudy and Uno, Shumpei},
  url={http://dx.doi.org/10.1103/PhysRevA.106.012423},
  DOI={10.1103/physreva.106.012423},
  journal={Physical Review A},
  volume={106},
  number={1},
  pages={012423},
  year={2022},
  publisher={APS}
}

@misc{readout_arxiv_2026,
      title={Tensor network characterization and mitigation of readout errors}, 
      author={Yuchen Guo and Shuo Yang},
      year={2026},
      eprint={2606.25974},
      archivePrefix={arXiv},
      primaryClass={quant-ph},
      url={https://arxiv.org/abs/2606.25974}, 
}

@article{linden2025use,
  title={How to use arbitrary measuring devices to perform almost-perfect measurements},
  author={Linden, Noah and Skrzypczyk, Paul},
  url={https://doi.org/10.1103/8tph-mc2p},
  DOI={10.1103/8tph-mc2p},
  journal={Physical Review A},
  volume={112},
  number={2},
  pages={022405},
  year={2025},
  publisher={APS}
}

@article{ouyang2024robust,
  title={Robust projective measurements through measuring code-inspired observables},
  url={http://dx.doi.org/10.1038/s41534-024-00904-y},
  DOI={10.1038/s41534-024-00904-y},
  author={Ouyang, Yingkai},
  journal={npj Quantum Information},
  volume={10},
  number={1},
  pages={104},
  year={2024},
  publisher={Nature Publishing Group UK London}
}

@article{byrne2026reducing,
  title={Reducing measurement error with adaptivity},
  author={Byrne, James and Linden, Noah and Skrzypczyk, Paul},
  journal={arXiv preprint arXiv:2606.21283},
  url={https://arxiv.org/abs/2606.21283},
  year={2026}
}

@article{cosco2025bayesian,
  title={Bayesian mitigation of measurement errors in multiqubit experiments},
  author={Cosco, Francesco and Plastina, F and Lo Gullo, N},
  url={http://dx.doi.org/10.1103/d65d-x8lt},
  DOI={10.1103/d65d-x8lt},
  journal={Physical Review A},
  volume={112},
  number={4},
  pages={042621},
  year={2025},
  publisher={APS}
}

@article{smith2021qubit,
  title={Qubit readout error mitigation with bit-flip averaging},
  author={Smith, Alistair WR and Khosla, Kiran E and Self, Chris N and Kim, MS},
  url={http://dx.doi.org/10.1126/sciadv.abi8009},
  DOI={10.1126/sciadv.abi8009},
  journal={Science advances},
  volume={7},
  number={47},
  pages={eabi8009},
  year={2021},
  publisher={American Association for the Advancement of Science}
  }

@article{maciejewski2020mitigation,
   title={Mitigation of readout noise in near-term quantum devices by classical post-processing based on detector tomography},
   volume={4},
   ISSN={2521-327X},
   url={http://dx.doi.org/10.22331/q-2020-04-24-257},
   DOI={10.22331/q-2020-04-24-257},
   journal={Quantum},
   publisher={Verein zur Forderung des Open Access Publizierens in den Quantenwissenschaften},
   author={Maciejewski, Filip B. and Zimborás, Zoltán and Oszmaniec, Michał},
   year={2020},
   month=Apr, pages={257} }

@article{bravyi2021mitigating,
  title={Mitigating measurement errors in multiqubit experiments},
  author={Bravyi, Sergey and Sheldon, Sarah and Kandala, Abhinav and Mckay, David C and Gambetta, Jay M},
  url={http://dx.doi.org/10.1103/PhysRevA.103.042605},
   DOI={10.1103/physreva.103.042605},
  journal={Physical Review A},
  volume={103},
  number={4},
  pages={042605},
  year={2021},
  publisher={APS}
}

@article{Nation2021,
  title={Scalable mitigation of measurement errors on quantum computers},
  author={Nation, Paul D and Kang, Hwajung and Sundaresan, Neereja and Gambetta, Jay M},
  url={http://dx.doi.org/10.1103/PRXQuantum.2.040326},
  DOI={10.1103/prxquantum.2.040326},
  journal={PRX Quantum},
  volume={2},
  number={4},
  pages={040326},
  year={2021},
  publisher={APS}
}

\end{document}